\documentclass[twocolumn,prb,aps,floatfix,longbibliography,superscriptaddress]{revtex4-1}

\usepackage{color}
\usepackage{graphicx}
\usepackage{siunitx}
\usepackage[color=green!60,textsize=small]{todonotes}
\usepackage{physics}
\usepackage{amsthm}
\usepackage{amsmath}
\usepackage{amssymb}
\usepackage{enumerate}
\usepackage{placeins}
\usepackage{booktabs}
\usepackage{dsfont}
\usepackage{hyperref}

\renewcommand{\vec}[1]{\boldsymbol{#1}}

\newcommand\mcm{\operatorname{MCM41}}
\newcommand{\kB}{k_{\mathrm{B}}}

\AtBeginDocument{%
\heavyrulewidth=.08em
\lightrulewidth=.05em
\cmidrulewidth=.03em
\belowrulesep=.65ex
\belowbottomsep=0pt
\aboverulesep=.4ex
\abovetopsep=0pt
\cmidrulesep=\doublerulesep
\cmidrulekern=.5em
\defaultaddspace=.5em
}

\begin{document}

\title{Dimensional Reduction of Helium-4 Inside Argon Plated MCM-41 Nanopores}

\author{Nathan S. Nichols}
\affiliation{Department of Physics, University of Vermont, Burlington, VT 05405, USA}
\affiliation{Materials Science Program, University of Vermont, Burlington, VT
05404,USA}

\author{Timothy R. Prisk}
\affiliation{Center for Neutron Research, National Institute of Standards and Technology, Gaithersburg, MD 20899-6100, USA}

\author{Garfield Warren}
\affiliation{Department of Physics, Indiana University, Bloomington, IN 47408, USA}

\author{Paul Sokol}
\affiliation{Department of Physics, Indiana University, Bloomington, IN 47408, USA}

\author{Adrian Del Maestro}
\affiliation{Department of Physics and Astronomy, University of Tennessee, Knoxville, TN 37996, USA}
\affiliation{Min H. Kao Department of Electrical Engineering and Computer Science, University of Tennessee, Knoxville, TN 37996, USA}
\affiliation{Department of Physics, University of Vermont, Burlington, VT 05405, USA}

\begin{abstract}
    The angstrom-scale coherence length describing the superfluid wavefunction of $^4$He at low temperatures has prevented its preparation in a truly one-dimensional geometry.  Mesoporous ordered silica-based structures, such as the molecular sieve MCM-41, offer a promising avenue towards physical confinement, but the minimal pore diameters that can be chemically synthesized have proven to be too large to reach the quasi-one-dimensional limit.  We present an active nano-engineering approach to this problem by pre-plating MCM-41 with a single, well controlled layer of Ar gas before filling the pores with helium.  The structure inside the pore is investigated via experimental adsorption isotherms and neutron scattering measurements that are in agreement with large scale quantum Monte Carlo simulations. The results demonstrate angstrom and kelvin scale tunability of the effective confinement potential experienced by $^4$He atoms inside the MCM-41, with the Ar layer reducing the diameter of the confining media into a regime where a number of solid layers surround a one-dimensional quantum liquid.
\end{abstract}

\maketitle


\section{Introduction}

The spatial dimension of a quantum many-body system can be systematically controlled by applying confinement on a scale smaller than the length characterizing coherence of the wavefunction. In this manner, one-dimensional (1D) phenomena have been explored in carbon nanotubes \cite{Bockrath:1999ww, Yao:1999hj, Ishii:2003my}, and low-density electronic quantum wires \cite{Auslaender:2005gb, Jompol:2009fe,Laroche:2014il, Blumenstein:2011gh}, where electron-beam lithography can achieve transverse confinement in the \SIrange{10}{100}{\nano\meter} range required to be smaller than the inverse Fermi wavevector. In ultra-cold atomic systems, laser trapping can produce confinement on the scale of the thermal de Broglie wavelength \cite{Monien:1998hc, Greiner:2001cma, Paredes:2004fp, Kinoshita:2005iz, Haller:2010fb, Boris:2016iz, Yang:2017fm}.  At higher densities, coherent quantum phenomena in the elemental superfluid $^4$He is characterized by a length scale $\xi(T) \approx \SI{1}{\nano\meter}$ below the superfluid transition temperature $T < T_{\lambda} \simeq \SI{2.12}{\kelvin}$ and engineering 1D confinement at this sub-nanometer scale has turned out to be a challenging task.

Current approaches to the physical confinement of superfluid helium fall into two categories: nanofabrication and chemical synthesis. In the first, electron beams have been employed to carve single short ($L < \SI{50}{\nano\meter}$) cylindrical pores with radii $R = \SIrange{3}{100}{\nano\meter}$ \cite{Savard:2009eq,Savard:2011fe,Duc:2015pd} while heavy-ion bombardment of polymer foils can create longer ($L = \SIrange{1}{100}{\micro\meter}$) and wider ($R = \SIrange{15}{200}{\nano\meter}$) channels \cite{Velasco:2012de, Velasco:2014gx,Botimer:2016pl}. In both cases, the measured hydrodynamics of the confined superfluid indicates deviations from bulk three dimensional pressure driven flow, providing evidence for a crossover towards the 1D limit.  The second approach invokes chemistry to synthesize silicates such as MCM-41 (Mobil Composition of Matter No.~41) \cite{Kresge:1992kd} and FSM-16 (Folded Sheet Material) \cite{Inagaki:1996fw} that consist of regular networks of hexagonal or cylindrical pores.  When filled with helium, they are amenable to bulk probes at low temperature and have provided a large body of evidence on the effects of enhanced thermal and quantum fluctuations on the superfluid state \cite{Wada:2001jb,Wada:2005uc,Ikegami:2005ec,Toda:2007cv, Taniguchi:2011bx, Taniguchi:2013us, Yager:2013cv, Demura:2015hq, Demura:2017gy,Taniguchi:2018ip,Prisk:2013cu,Bryan:2017hb, Bryan:2018vb, Bossy:2019qd, Taniguchi:2020ln,Huan:2020ya}. However, the radii of the pores in these materials is ultimately set by the specific reaction route and is not continuously tunable,  with the smallest possible diameter being on the order of \SI{2}{\nano\meter}.  Quantum Monte Carlo simulations of confined $^4$He have indicated that sub-\si{\nano\meter} radii might be required to observe truly 1D behavior \cite{DelMaestro:2011dh, DelMaestro:2012ba,  Kulchytskyy:2013dh, Markic:2015bu, Markic:2018bw,Markic:2020ip} and thus a systematic approach to reducing the size of nanopores is desirable to test these predictions. 

In this paper, we introduce a proposed solution which employs pre-plating MCM-41 nanopores with a single adsorbed layer of argon gas \cite{Muroyama:2008ky,Kilburn:2008kl}, thereby allowing tunability of both the effective pore radius seen by helium atoms and the strength of the confinement potential.  We combine experimental results employing N$_2$ and $^4$He adsorption isotherms with large scale quantum Monte Carlo simulations to explore the atomic-scale structure within the pores and identify a promising region where the density of a central core of helium atoms may be manipulated upon filling.

A single nanopore of Ar pre-plated MCM-41 is modelled by constructing an
effective confinement potential consisting of a superposition of Lennard-Jones
terms for He interacting with atoms in the porous material and a single
cylindrical shell of argon.  The resulting potential is tunable, both in terms
of the location and depth of its wall-proximate minima.  Its specific form can
be matched to the microscopic geometry by extracting the width and density of
the argon layer via a Brunauer-Emmett-Teller (BET) analysis \cite{Sing:1985rg}
of experimental adsorption isotherms.  Once fixed, this potential is employed
in a grand canonical quantum simulation of helium inside the nanopore where the density can be tuned by modifying the chemical potential (corresponding to the pressure of an external reservoir).  As the pressure is increased, a series of concentric cylindrical layers form with the outer-most shells near the Ar exhibiting solid-like behavior.  As the pressure approaches that of saturated vapor, the pores become fully filled and exhibit a central column of helium which may realize the desired 1D behavior.  The existence of a central column is not a generic effect, but is instead a result of the ratio of the pore radius and the location of the mimima of the He - He interaction potential being close to an integer value.  Careful analysis of simulation data allows for the determination of the relation between the linear density (number of atoms) in the pore center and the external pressure showing a narrow window, \SIrange{0.0718}{1.635}{\pascal} at $T = \SI{1.6}{\kelvin}$, where a compressible 1D liquid can be expected inside the pre-plated nanopores.

The remainder of this paper is organized as follows.  We first describe the experimental synthesis of MCM-41 and our pre-plating procedure followed by a characterization of the nanoporous materials via adsorption isotherms, elastic neutron scattering, and inelastic neutron scattering.  The results allow us to extract material parameters that are essential in the construction of the model pre-plated MCM-41 confinement potential that is employed in a quantum Monte Carlo methodology based on path integrals. We next present the results of numerical simulations detailing the structure inside the pore as the external pressure is increased. We conclude with an analysis of the resulting layer formation and discuss implications for the discovery of a tunable 1D liquid in this geometry.

\section{Experimental Results}
\label{sec:experimental_results}

\subsection{Sample Characterization}
MCM-41 is a mesoporous material with a hierarchical structure produced using a surfactant templating technique.  The surfactants used form rod-like micelles that order in a hexagonal array.  The pores of this material, after removal of the surfactant template, are monodisperse, unidirectional, and have a regular 2D hexagonal structure.  The typical aspect ratio of the pores is $\sim 1000:1$.  

Our sample was obtained from Sigma-Aldrich\cite{mcm41SA:2008} and was characterized using X-ray powder diffraction and N$_2$ gas adsorption isotherm measurements. The X-ray diffraction data indicated that the sample consisted of a single phase with pores arranged on a hexagonal lattice with a lattice constant of \SI{4.7}{\nano\meter}.  A Brunauer-Emmett-Teller (BET) analysis \cite{Brunauer:1938pz} of the N$_2$ isotherm gave a surface area of \SI{915}{\meter^2\per\gram}.  The pore diameter size distribution was calculated using the Kruk-Jaroniec-Sayari method \cite{Jaroniec:1999mi} and was found to be Gaussian with a mean value of \SI{3.0}{\nano\meter} and a full-width at half-maximum of \SI{0.3}{\nano\meter}. 

\begin{figure}[ht]
    \centering
    \includegraphics[width=\columnwidth]{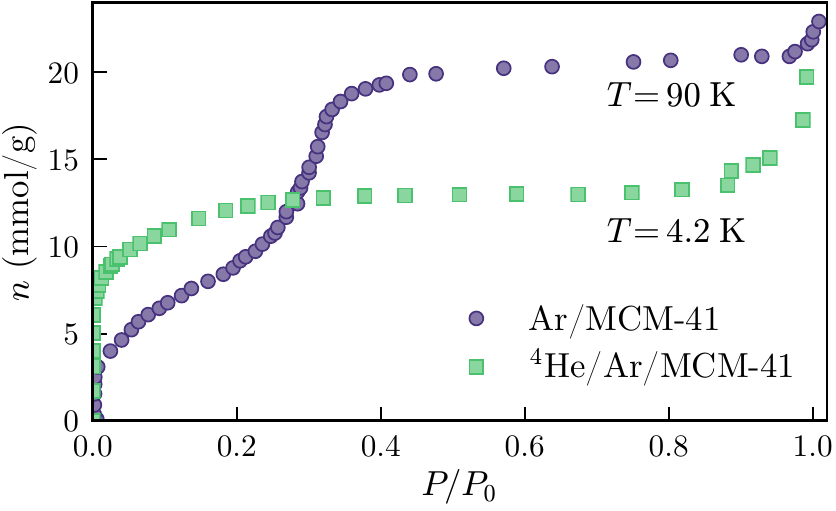}
    \caption{Experimental adsorption isotherms of our MCM-41 sample collected under two different conditions.  The purple circles indicate the amount of argon adsorbed on the untreated material as a function of pressure at a fixed temperature of 90 K.  Green squares illustrate the adsorption behavior of $^4$He at 4.2 K on to MCM-41 already pre-plated with a monolayer of Ar gas.  Here $P_0$ is the bulk equilibrium vapor pressure of Ar ($^4$He) for the purple circles (green squares).}
    \label{fig:4Heisotherm}
\end{figure}

Adsorption isotherms were also carried out with research grade Ar gas to determine the monolayer coverage for the pores at \SI{90}{\kelvin}. The results are shown in Figure~\ref{fig:4Heisotherm}. A BET analysis of the isotherm yielded a monolayer coverage of \SI{8.994}{\milli\mol\per\gram}. This monolayer coverage, when combined with the measured surface area, yields an aerial coverage of $\SI{0.59}{\angstrom^{-2}}$ and, using the van der Waals radius for Ar, a monolayer density of $n_{\rm Ar} = \SI{0.017}{\angstrom^{-3}}$.

\subsection{$^4$He Isotherms}
\label{subsec:4HeIsotherms}

We also carried out $^4$He isotherms on MCM-41 preplated with a single monolayer of Ar.  The Ar pre-plating was carried out at \SI{90}{\kelvin} and then the sample was slowly cooled to \SI{4.2}{\kelvin} over the course of several hours.  $^4$He isotherms were then carried out at \SI{4.2}{\kelvin} using standard volumetric techniques.  The results are shown in Figure~\ref{fig:4Heisotherm}.  The $^4$He initially adsorbed is strongly bound to the surface resulting in zero pressure rise until $\sim \SI{7.5}{\milli\mol\per\gram}$ has been adsorbed.  There is a small region between $\sim\SI{7.5}{\milli\mol\per\gram}$ and \SI{13}{\milli\mol\per\gram}  where the pressure increases.  Once a  filling of 13 mmol/g has been reached no additional helium is adsorbed into the pores until the pressure is close to the bulk vapor pressure.  Once P/P$_0$ is greater than $\sim 0.9$, $^4$He capillary condenses between the MCM-41 grains.

\subsection{Neutron Scattering}

Neutron scattering studies of $^4$He in Ar preplated MCM-41 were performed to
identify the phase (mobile versus immobile) of the adsorbed helium. These
measurements were carried out using the Disc Chopper Spectrometer (DCS) at the NIST Center for Neutron Research \cite{Copley:2003dc}. This instrument is a direct
geometry time-of-flight chopper spectrometer which views a cold moderator. An
incident wavelength of \SI{2.5}{\angstrom^{-1}} was used for these measurements.  A top-loading liquid helium cryostat with aluminum tails, commonly referred to as an ``orange" cryostat, was used to obtain the low temperatures examined in this study. The sample cell was a cylindrical aluminum can of outer diameter \SI{1.5}{\centi\meter}, a height of \SI{6}{\centi\meter}, and a wall thickness of \SI{1}{\milli\meter}.  The cell contained \SI{6.13}{\gram} of MCM-41 in the form of cylindrical pellets \SI{1}{\centi\meter} thick separated by cadmium spacers to reduce multiple scattering.  Gas was loaded to the sample \textit{in situ} from an external gas handling system. Measurements were carried out at a temperature of \SI{1.6}{\kelvin}. Standard data reduction routines \cite{Azuah:2009cs} were used to convert the observed scattering to the dynamic structure factor $S(Q,E)$.
\begin{figure}[ht]
    \centering
    \includegraphics[width=\columnwidth]{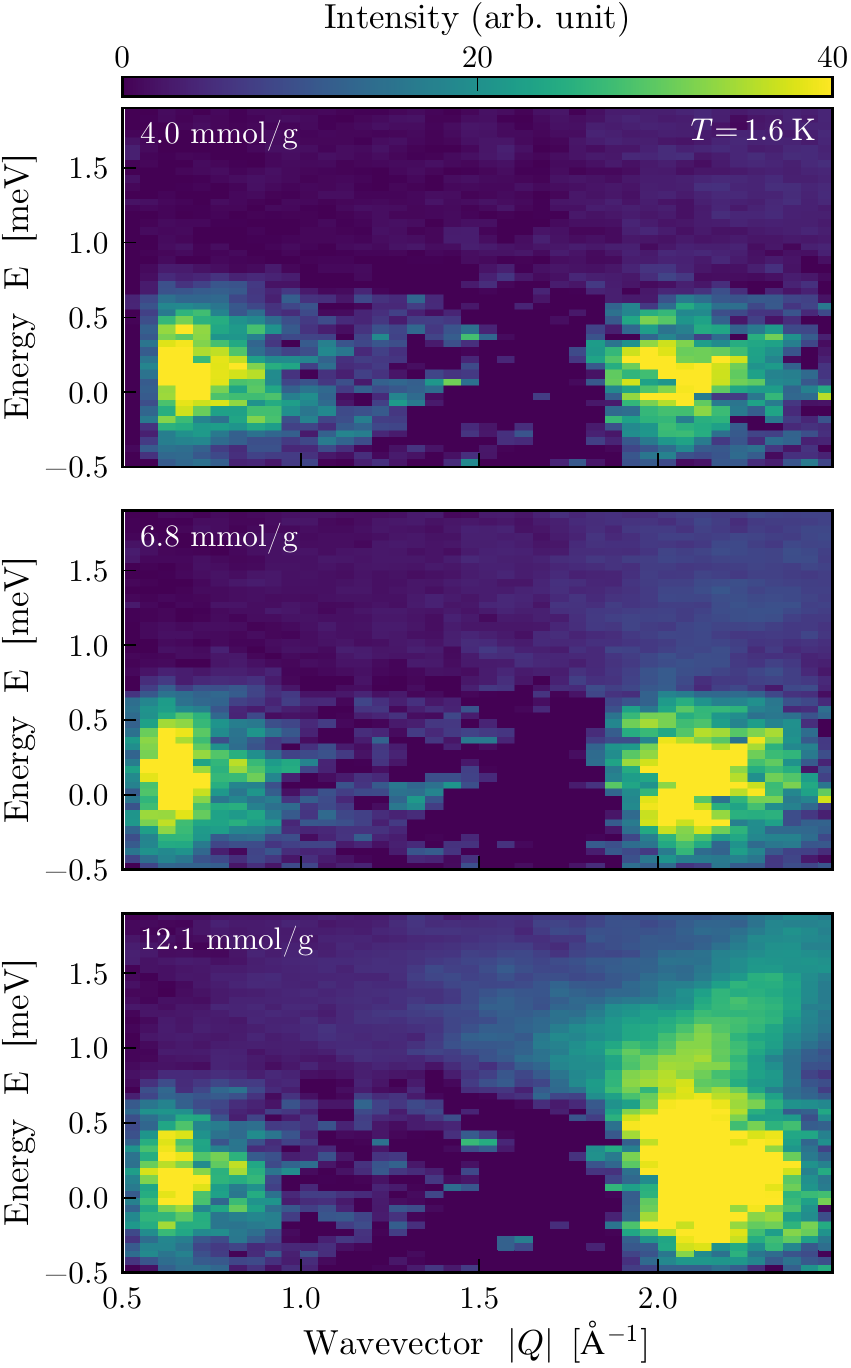}
    \caption{The dynamic structure factor $S(Q,E)$ of $^4$He inside MCM-41 that has been pre-plated with a single layer of Ar at \SI{1.6}{\kelvin}. The scattering from the cell, MCM-41, and the Ar layer have been subtracted.  The panels from top to bottom show results for pore fillings \SIlist{4.0;6.8;12.1}{\mmol\per\gram} and correspond to a single layer, double layer, and a completely filled pore. The mostly elastic scattering in the top and middle panel demonstrate the quasi-two-dimensional solid-like behavior of the adsorbed helium near the argon layer, while the dispersing inelastic intensity emerging from $|Q| \simeq \SI{2}{\angstrom}$ supports the existence of a liquid supporting density-wave excitations at the center of the pore.}
    \label{fig:scattering}
\end{figure}

The sample was preplated with a monolayer of Ar prior to the adsorbing ${^4}$He.
The scattering from the cell, MCM-41, and Ar monolayer were treated as
background and subtracted from the results shown in Figure~\ref{fig:scattering}.  The three panels show scattering at three different fillings corresponding to (top) monolayer, (middle) bilayer, and (bottom) full pore.  For the purposes of this discussion, we are only interested in the information the scattering provides on the mobility of the
adsobed helium.  A more complete analysis of the scattering will be presented elsewhere.  For the monolayer and bilayer, only elastic scattering ($E=0$) is
observed.  The strong scattering at $\sim\SI{2.5}{\angstrom^{-1}}$ represents the first peak in the static structure factor $S(Q)$ for the adsorbed helium. The full pore measurement, in contrast, exhibits inelastic scattering consistent with mobile $^4$He atoms. 

 The lack of inelastic scattering for the monolayer and bilayer indicate that when ${^4}$He is initially adsorbed on the Ar plated MCM-41 pores it is strongly bound and immobile.  This is consistent with the predictions of the simulations (discussed below) which predict high density solid like layer formation for the first few layers of ${^4}$He adsorbed in the pores.  The appearance of both elastic scattering and inelastic scattering for the nearly full pores is consistent with the simulation results that predict mobile (low density) helium at the center of the pores surrounded by multiple solid layers.

\section{Model and Simulation Details}

The data obtained from the above experimental characterization of the Ar pre-plated MCM-41 nanopores can now be used to construct a theoretical model of the confinement geometry. We begin by simplifying the analysis to a single pore, as their center-to-center separation of \SI{4.7}{\nano\meter} means that atoms in different pores are essentially non-interacting.  The one-pore system can then be described by the $N$-body Hamiltonian:
\begin{equation}
H = -\frac{\hbar^2}{2m}\sum_{i=1}^N \nabla_i^2 + \sum_{i=1}^N U(\vec{r}_i) + \frac{1}{2}\sum_{i,j} V(\vec{r}_i-\vec{r}_j)
\label{eq:Ham}
\end{equation}
where $m$ is the mass of a ${^4}$He atom located at position $\vec{r}_i = (x_i,y_i,z_i)$ confined inside a pre-plated nanopore by a single particle potential energy $U$ and interacting with other He atoms through $V$.  Both potential energy terms arise from induced dipole-dipole interactions, with the helium-helium interaction potential being known to high precision \cite{Aziz:1979hs, Przybytek:2010js, Cencek:2012iz}.  $U$ is more difficult to obtain and its estimation now proceeds by generalizing previous results for the confinement of helium inside a cylinder carved out of an infinite homogeneous medium \cite{Tjatjopoulos:1988ec,Stan:1999do, Zhang:2004cw, Rossi:2006cj}. 



\subsection{Pre-plated confinement potential}
\label{ssec:preplatePotential}

\begin{figure}[t]
    \includegraphics[width=\columnwidth]{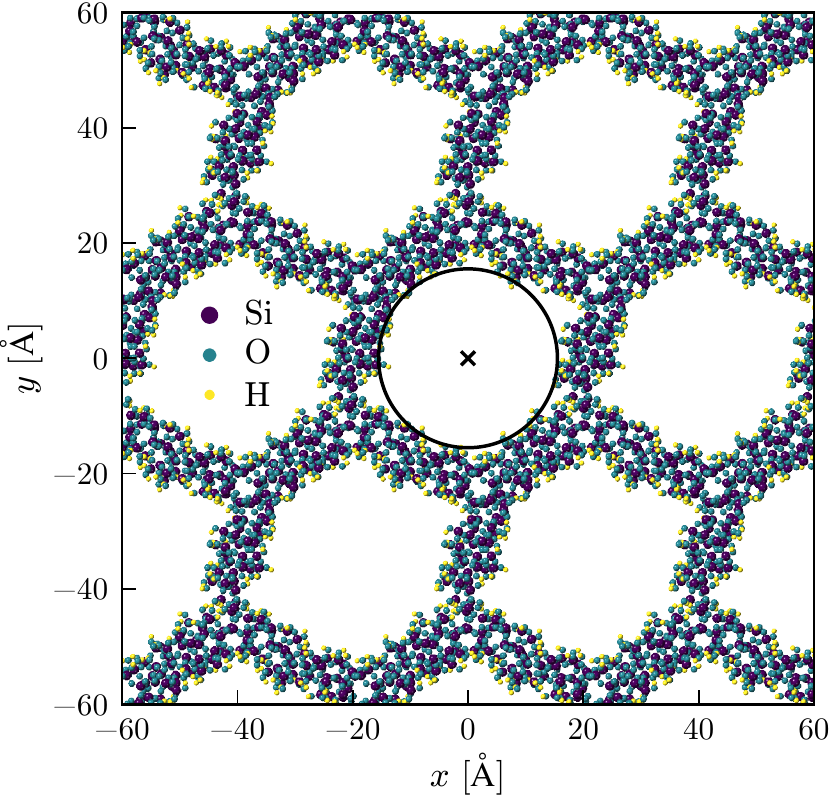}
    \caption{A projection of a MCM-41 supercell into the $xy$-plane showing the nearly cylindrical pores and positions of H, O, and Si atoms obtained from a density functional theory optimized structure employing the B3LYP/6-31G(d,p) basis set \cite{Ugliengo:2008ks}. The cross indicates the origin of the coordinate system while the circle is plotted at the determined pore radius $R=\SI{15.51}{\angstrom}$.}
    \label{fig:mcm41}
\end{figure}

\begin{figure}[t]
    \centering
    \includegraphics[width=\columnwidth]{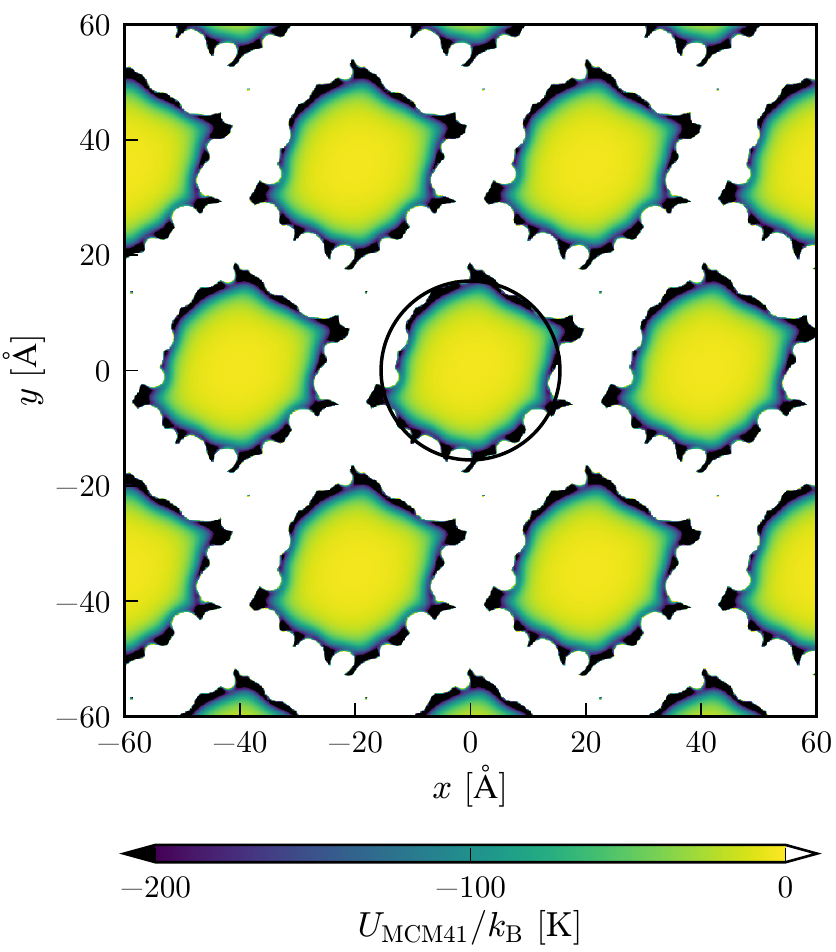}
    \caption{\label{fig:manybodyPotential} The effective manybody potential in units of $k_{\rm B}$ plotted at a single slice of the $xy$-plane at $z=\SI{0.0}{\angstrom}$ in the range \SI{-200}{\kelvin} to \SI{200}{\kelvin}. Features of the pore
    roughness can be seen here, with potential well depths approaching
    \SI{-500}{\kelvin} for this slice. Pre-plating material is expected to fill
    in the nooks and crannies of the pore wall.
    }
\end{figure}

We begin by considering the potential environment for a $^4$He atom inside a single pore.  Figure~\ref{fig:mcm41} shows a bird's-eye view of the structure of MCM-41 ($z$-axis points out of the page) obtained via density functional theory \cite{Ugliengo:2008ks}. 
The atomic coordinates have been shifted such that the origin $(0,0,0)$ is defined to be at the center of a pore (as indicated by the star).  The black circle describes a perfect cylinder with radius $R=\SI{15.51}{\angstrom}$ constructed to fit within the quasi-hexagonal pores.  This value is in agreement with the average pore radii of MCM-41 extracted from a BET analysis of experimental data. The resulting confinement potential for a helium atom at position $\vec{r}_i$ is then formed from the usual sum over Lennard-Jones pair-wise contributions:
\begin{equation}
    U_{\mcm}(\vec{r_i}) = 4\sum_j \varepsilon_{ij} \qty(\abs{\frac{\sigma_{ij}}{{\vec{r}_i-\vec{r}_j}}}^{12} 
        - \abs{\frac{\sigma_{ij}}{{\vec{r}_i-\vec{r}_j}}}^{6})
\label{eq:VHeMCM}
\end{equation}
where $\varepsilon_{ij}$ and $\sigma_{ij}$ are estimated with
Lorenz-Bertholot mixing rules \cite{Boda:2008yy} for two atomic species
$i$ and $j$:
\begin{equation}
\begin{aligned}
\varepsilon_{ij} &= \sqrt{\varepsilon_i \varepsilon_j} \\
    \sigma_{ij} &= \frac{\sigma_i + \sigma_j}{2}.
\end{aligned}
\label{eq:LBrules}
\end{equation}
The brute-force sum over $j$ can be extended to a large number of unit cells to
obtain convergence to some fixed numerical precision with details, including all Lennard-Jones parameters, described in Appendix~\ref{app:potential}. The result for a single slice at $z=\SI{0.0}{\angstrom}$ is shown in Figure~\ref{fig:manybodyPotential} where the potential has only been plotted in the range $-\SI{200}{\kelvin} \le U_{\mcm}(x,y,z=0)/k_{\rm B} \le \SI{200}{\kelvin}$. 
In the deep pockets near the pore walls indicated in the figure, the depth of
the well can drop to nearly \SI{-800}{\kelvin} for some values of $z$. 
We expect that upon pre-plating the MCM-41 with a light rare gas such as argon, these recesses will be completely filled via adsorption effects.  Thus the resulting potential seen by helium will be considerably smoother.  Motivated by this, we model the confinement potential felt by a helium atom at 
radius $r$ from the center of a pore using a radially symmetric effective potential inside a long uniform cylinder of radius $R$ carved inside a continuous media \cite{Zhang:2004cw}. The resulting potential is now a scalar function of $r$:
\begin{multline}
    U_{\rm cyl} (r;n,\varepsilon, \sigma, R) = \\ 
    \frac{\pi n \varepsilon\sigma^3}{3}\qty[\qty(\frac{\sigma}{R})^9 \mathop{u_9}\qty(\frac{r}{R}) - \qty(\frac{\sigma}{R})^3 \mathop{u_3}\qty(\frac{r}{R})]
    \label{eq:Vpore}
\end{multline}
with
\begin{multline*}
    u_9(x) = \frac{1}{240(1-x^2)^9} \bigl[ \\
    (1091 + 11156x^2 + 16434x^4 + 4052x^6 + 35x^8)E(x) \\
    - 8(1-x^2)(1+7x^2)(97+134x^2+25x^4)K(x)\bigr]
\end{multline*}
and
\begin{equation*}
    u_3(x) = \frac{2}{(1-x^2)^3} \qty[(7+x^2)E(x) - 4(1-x^2)K(x)]
\end{equation*}
where $n$ is the density of the media, $\varepsilon$ is the strength of the interaction, $\sigma$ is the hard core distance, $R$ is the pore radius, and $K(x)$ and $E(x)$ are the complete elliptic integrals of the first and second kind.

The values of $\sigma$, $n \varepsilon$, and $R$ in Eq.~(\ref{eq:Vpore}) can be extracted through a non-linear least squares fitting procedure of $U_{\mcm}$ to $U_{\rm cyl}$ as described in Appendix~\ref{app:potential}. The resulting quantities are reported in Table~\ref{tab:MCM41} for helium inside MCM-41.  
\begin{table}
    \renewcommand{\arraystretch}{1.5}
    \setlength\tabcolsep{12pt}
    \begin{tabular}{@{}lll@{}} 
        \toprule
        $R\; \qty[\si{\angstrom}]$ & $\sigma\; \qty[\si{\angstrom}]$ & $n \varepsilon/k_\mathrm{B}\; \qty[\si{\kelvin\angstrom^{-3}}]$ \\
        \midrule
        15.51  & 3.44 & 1.59  \\
        \bottomrule
    \end{tabular}
    \caption{\label{tab:MCM41} The effective Lennard-Jones parameters used in a cylindrical model of MCM-41 described by Eq.~(\ref{eq:Vpore}) with details in provided Appendix~\ref{app:potential}.}
\end{table}

With an effective potential described by Eq.~(\ref{eq:Vpore}) for helium inside MCM-41 we can now model the rare gas pre-plating by superimposing a continuous cylindrical shell with of width $w = R_{\rm out} - R_{\rm in}$ that yields additional confinement
\begin{align}
    U_\mathrm{shell}(r) &= U_{\rm cyl}(r;n_{\rm Ar},\varepsilon_{\rm Ar-He}, \sigma_{\rm Ar-He},  R_\mathrm{in})  \nonumber \\
                        &- U_{\rm cyl}(r;n_{\rm Ar},\varepsilon_{\rm Ar-He},
                        \sigma_{\rm Ar-He},R_\mathrm{out})
    \label{eq:Vshell}
\end{align}
where $R_{\rm out} = \SI{15.51}{\angstrom}$ is the outer radius computed from
$U_{\rm cyl}$ and $R_{\rm in} = \SI{11.75}{\angstrom}$  is the inner radius
using the van der Waals diameter of Ar. The mixed $^4$He-Ar  
Lennard-Jones parameters can be computed from the values in Table~\ref{tab:LJ1}.
\begin{table}
    \renewcommand{\arraystretch}{1.5}
    \setlength\tabcolsep{12pt}
    \begin{tabular}{@{}lll@{}} 
        \toprule
        Atom & $\sigma\; \qty[\si{\angstrom}]$ & $\varepsilon/k_\mathrm{B}\; \qty[\si{\kelvin}]$ \\
        \midrule
        He     & 2.640 & 10.9    \\
        Ar     & 3.405 & 119.8   \\
        \bottomrule
    \end{tabular}
    \caption{\label{tab:LJ1} Lennard-Jones parameters \cite{1964:HirschfelderBook} used in the evaluation of the pre-plating layer potential defined in Eq.~(\ref{eq:Vshell}).}
\end{table}

\begin{figure}[t]
    \centering
    \includegraphics[width=\columnwidth]{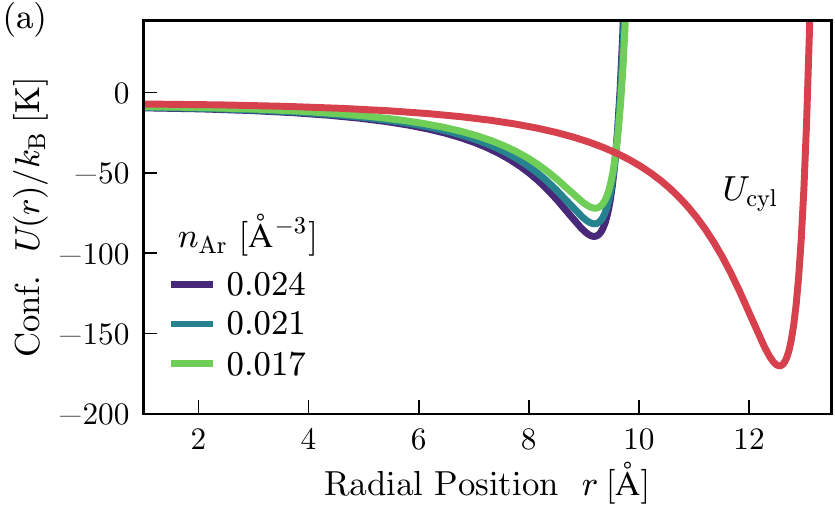}\\[1.0ex]
        \includegraphics[width=\columnwidth]{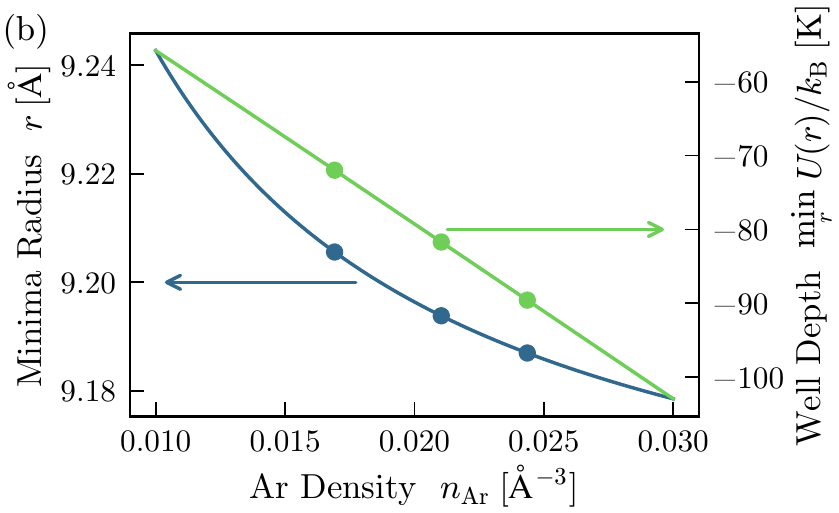}
    \caption{The top panel (a) shows helium interacting with a shell of argon
    at various densities and the mesoporous silica, MCM-41. The potentials can
    be calculated using the parameters shown in Table~\ref{tab:LJ1} and
    Table~\ref{tab:MCM41} with Eq.~(\ref{eq:Vpore}) and Eq.~(\ref{eq:Vshell}).
    The bottom panel (b) shows how varying the density affects the minimum
    location (left axis) and potential well depth (right axis).}
    \label{fig:effectivePotential}
\end{figure}

The full interaction potential $U(r) = U_{\rm cyl}(r) + U_{\rm shell}(r)$ for helium inside the Ar pre-plated MCM-41 system is shown in Figure~\ref{fig:effectivePotential}(a) for three different densities of the argon layer corresponding to $n_{\rm Ar} = \SI{0.017}{\angstrom^{-3}}$ determined from the experimental BET analysis of an Ar/MCM-41 isotherm in Section~\ref{subsec:4HeIsotherms}, $n_{\rm Ar} = \SI{0.021}{\angstrom^{-3}}$ for Ar liquid at boiling point, and $n_{\rm Ar} = \SI{0.024}{\angstrom^{-3}}$ for solid Ar at its triple point \cite{VanWitzenburg:1967}.  The bare interaction potential for helium with MCM-41, $U_{\rm cyl}$, is shown for comparison.
We observe that the density of the plated layer affects both the depth of the well and the location of the minimum, with this relationship quantified in Figure~\ref{fig:effectivePotential}(b).    Modifications of the density of the pre-plating layer provide substantial tuneability of the scale of the confinement potential seen by helium atoms while only producing a sub-\si{\angstrom} modification of its effective radius.  This indicates that substantially different levels of confinement can be produced by modifying the species of rare gas when pre-plating. With an estimate of the environment inside a single Ar preplated MCM-41 nanopore experienced by a single ${^4}$He atom, we now briefly describe the technical details of a quantum simulation of confined liquid helium at low temperature.

\subsection{Quantum Monte Carlo}

A system of $N$ helium atoms described by Eq.~(\ref{eq:Ham}) inside the pore can be simulated using a Monte Carlo technique that exploits the path integral representation to map the quantum system in $D=3$ spatial dimensions to an effective classical one in $D+1=4$ that can be efficiently sampled using the Metropolis-Hastings algorithm \cite{Ceperley:1995gr}.  A grand canonical worm algorithm suitable for bosons in the spatial continuum introduced by Boninsegni, Prokof'ev and Svistunov \cite{Boninsegni:2006ed,Boninsegni:2006gc} provides access to finite temperature observables:  $\expval{\mathcal{O}} \propto \mathrm{Tr}\, \mathcal{O} \, e^{-H/k_{\rm B} T}$ where $k_{\rm B}$ is the Boltzmann constant, for systems composed of a few thousand ${^4}$He atoms. 

All results presented herein utilized our open source path integral quantum
Monte Carlo code in the grand canonical ensemble (access details in Ref.~[\onlinecite{delmaestroSVN}]) and all code, scripts, and data used in analysis and plotting are available online \cite{repo}.  

We considered pores of lengths
$L=\SIrange{25}{100}{\angstrom}$ in order to understand any finite size effects (Appendix~\ref{ssec:finite_size_scaling}), and
focused on chemical potentials in the range $\mu/\kB = \SIrange{-100}{0}{\kelvin}$
at two experimentally studied temperatures: $T=\SI{4.2}{\kelvin}$ and
\SI{1.6}{\kelvin}.  The imaginary time step was fixed at $k_{\rm B}\tau =
\SI{0.004}{\kelvin^{-1}}$ after comparing systematic Trotter and statistical errors for this value (see Appendix~\ref{ssec:tau_scaling}). 

\begin{figure}[t]
\begin{center}
\includegraphics[width=1.0\columnwidth]{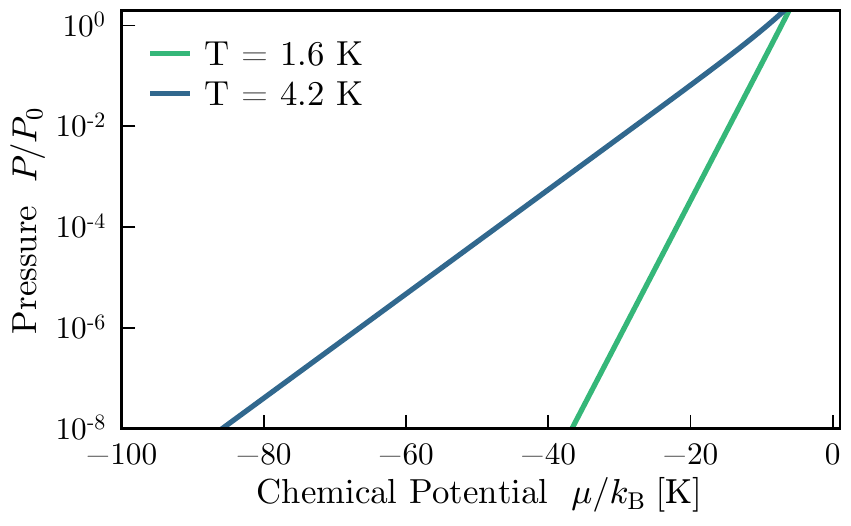}
\end{center}
\caption{Pressure as a function of chemical potential for bulk $^4$He.  The relationship is computed via the second order virial expansion (Eq.~\ref{eq:virialP}) using the tabulated temperature dependence of the thermodynamic properties of helium at low temperature \cite{Donnelly:1998ge}.}
\label{fig:Pvsmu}
\end{figure}

In order to make connections between simulations and experiment we can convert
the chemical potential (simulation tuning parameter) to pressure (experimental
control knob). This was achieved by employing the virial equation of state up
to second order using the known temperature dependence of the second coefficient $B_2(T)$ for bulk $^4$He at saturated vapor pressure \cite{Donnelly:1998ge} which yields
\begin{equation}
    P \simeq \frac{k_{\rm B}T}{\Lambda^3(T)}\mathrm{e}^{\mu/k_{\rm B}T} \qty[1 - B_2(T)\frac{\mathrm{e}^{\mu/k_{\rm B}T}}{\Lambda^3(T)}]
    \label{eq:virialP}
\end{equation}
where $\Lambda(T)=h/{\sqrt{2\pi m k_{\rm B}T}}$ is the thermal de Broglie wavelength. Eq.~(\ref{eq:virialP}) is shown in Figure~\ref{fig:Pvsmu} in units of the saturated vapor pressure of bulk helium, $P_0$, and is used throughout this work to convert between chemical potential and pressure.
%
%
The pure exponential dependence holds over nearly the entire range of chemical potentials with deviations appearing when $P \simeq P_0$.  A more accurate \emph{ab initio} estimation of the pressure inside the pore can be obtained via quantum Monte Carlo \cite{Ceperley:1995gr}. However, the effective phase separation inside the pore (see Section~\ref{sec:simulation_results}) makes statistical convergence of results difficult to obtain and we thus use the bulk relationship here. This simplification could introduce uncertainties when comparing with experimental results manifested as a pressure offset that can be overcome by fixing either the onset (initial) or saturation (maximal) filling of the pores.

\section{Simulation Results}
\label{sec:simulation_results}

All results presented in this section for $^4$He inside Ar plated MCM-41 have a fixed pore length of $L = \SI{50}{\angstrom}$, pore radius $R = \SI{15.51}{\angstrom}$, and
employ $U(\vec{r})$ described in Section~\ref{ssec:preplatePotential} in the grand canonical ensemble yielding $N \approx 600$ helium atoms for the largest chemical potentials studied. 

\begin{figure}[t]
    \centering
    \includegraphics[width=\columnwidth]{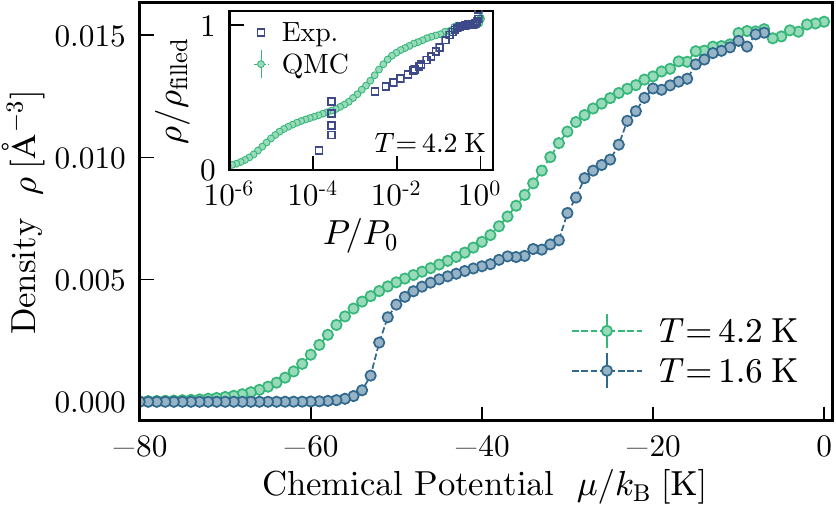}
    \caption{Adsorption isotherms showing the number density $\rho$ inside a MCM-41 nanopore as a function of the chemical potential $\mu$.  Step-like features indicate the onset of layer formation.  The inset shows a comparison between quantum Monte Carlo (filled circles) simulations and experimental results (open squares) at $T = \SI{4.2}{\kelvin}$ where the vertical scale has been normalized to a density corresponding to a completely filled pore. Differences between the numerical and experimental isotherm can be attributed the use of a smooth effective potential $U(r)$ in the quantum Monte Carlo.}
\label{fig:filling_the_pore}
\end{figure}

A separate simulation was performed for each chemical potential $\mu/\kB$ in the range \SIrange{-100}{-7}{\kelvin} at $T = \SI{1.6}{\kelvin}$ and 
\SIrange{-100}{0}{\kelvin} at $T = \SI{4.2}{\kelvin}$ in steps of \SI{1}{\kelvin}.  The number density $\rho = N/V$ where $V = \pi R^2 L$ is the volume of the pore, was computed with the results shown in Figure~\ref{fig:filling_the_pore}.
The adsorption isotherms demonstrate that the pore remains empty until a temperature-dependent critical chemical potential is reached and atoms start to enter the pore.  The density increases with increasing $\mu$ with step-like features indicative of layer formation consistent with previous simulation studies\cite{Boninsegni:2001jy, Rossi:2005hp, DelMaestro:2011dh, DelMaestro:2012ba,  Kulchytskyy:2013dh, Pollet:2014fb, Markic:2015bu}. We note that even as $\mu \to 0$ the density inside the pore, $\rho_{\rm filled}$, remains smaller than that of bulk helium at saturated vapor pressure \SI{0.019}{\angstrom^{-3}} for $T = \SI{4.2}{\kelvin}$ and $\SI{0.022}{\angstrom^{-3}}$ for $T = \SI{1.6}{\kelvin}$ \cite{Donnelly:1998ge}. The origin of this behavior can be investigated through a closer examination of the structure inside the pore as measured by a radial density:
\begin{equation}
    \rho_{\rm rad}(r) = \left \langle \sum_{i=1}^{N} \delta\qty(\abs{\vec{r}_i} -r) \right \rangle
\label{eq:radial_density}
\end{equation}
where $\vec{r}_i$ are the locations of the $^4$He atoms and $\langle \dots \rangle$ indicates a Monte Carlo average where it is noted that $N$ is an instantaneous (configuration-dependent) quantity in the grand canonical simulation.

\begin{figure}[t]
    \centering
    \includegraphics[width=\columnwidth]{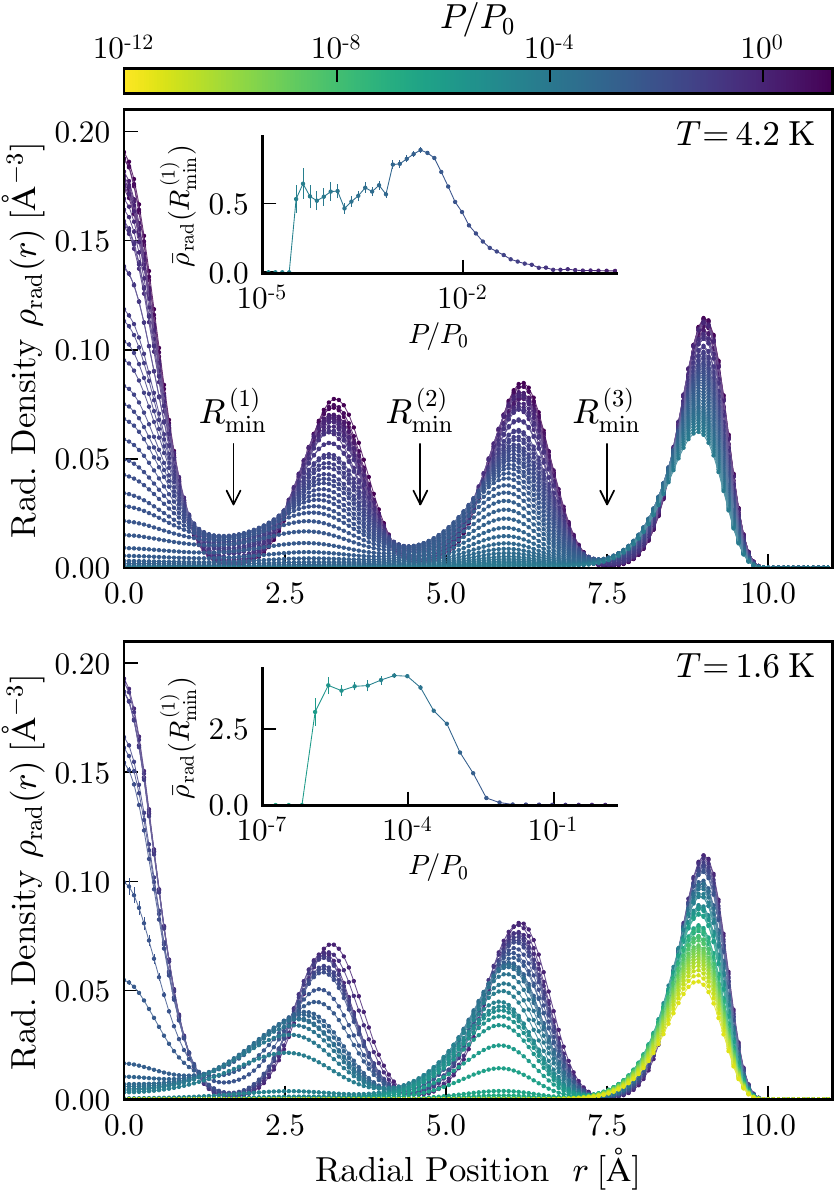}
    \caption{The radial density inside the Ar preplated MCM-41 for two temperatures computed via quantum Monte Carlo simulations.  The curves in the main panels correspond to different chemical potentials between $\mu/\kB = \SIrange{-50}{0}{\kelvin}$ which have been converted to pressures using Eq.~(\ref{eq:virialP}) to calibrate the displayed colorbar. The peaks show the buildup of concentric quasi-2D layers of helium and the existence of a quasi-1D core at pressures $P > 0.01P_0$ corresponding to $\mu/\kB > \SI{-17}{\kelvin}$. The insets detail the relative density $\rho_{\rm rad}(R_{\rm min}^{(1)})/\rho_{\rm rad}(0)$ at $r = R_{\rm min}^{(1)}$ corresponding to the position of the first minima as a function of pressure. The location of the minima are independent of pressure for $P>10^{-2}P_0$.}
    \label{fig:radialDensity}
\end{figure}

The resulting radial density is plotted as a function of radial position (distance from the center of the pore) in Figure~\ref{fig:radialDensity} for $\mu/\kB > \SI{-50}{\kelvin}$, where the chemical potential has been converted to pressure via Eq.~(\ref{eq:virialP}).
The onset of well-defined peaks in the radial density correspond to the steps in Figure~\ref{fig:filling_the_pore}.  At low pressure, a quasi-solid layer of helium forms near the hard wall created by the pre-plated argon shell. As the pressure is increased, a sequence of concentric quasi-2D shells form with near vacuum between them. The magnitude of the number density of the shells places them all within the quasi-solid regime for $^4$He.  

When $P$ is increased beyond $10^{-3}P_0$ ($\mu/\kB \gtrsim \SI{-17}{K}$), $^4$He atoms begin to fill an inner core.  The insets of Fig.~\ref{fig:radialDensity} show the density at the location of the first minimum normalized by the radial density at the center of the pore. While there is an intermediate range of pressures where atoms can move freely between the center of the pore and first shell, the density at the minima begins to drop precipitously at higher pressures.  For $P > 10^{-2}P_0$ the density between the inner core and first shell is vanishingly small and particle exchanges are strongly suppressed indicating that the core region is acting as a quasi-1D system.

We note that the existence of this central core is not a generic effect and represents the fact that in the geometry considered here, the ratio of the effective pore-radius (set by the MCM-41 and Ar pre-plating geometry) to the distance between shells (set by the helium interaction potential) is approaching an integer. In nanopores where the effective radii is different, the density of the central core can be vanishingly small \cite{DelMaestro:2011dh,DelMaestro:2012ba}.

We now focus on the details of the central core and shells which can be characterized by one- and two-dimensional densities defined by:
\begin{align}
\label{eq:rho1d}
\rho_{\rm 1D} &= 2 \pi \int_0^{R_{\rm min}^{(1)}} r\dd{r}\ \rho_{\rm rad}(r)  \\
\label{eq:rho2d}
\rho_{\rm 2D} &= \frac{1}{R_{\rm max}^{(j)}} \int_{R_{\rm min}^{(j)}}^{R_{\rm min}^{(j+1)}}r\dd{r}\ \rho_{\rm rad}(r)
\end{align}
where to avoid ambiguity, we have determined the locations of the $j^{th}$ minima ($R_{\rm min}^{(j)}$) and $j^{th}$ maxima ($R_{\rm max}^{(j)}$) at fixed $\mu/k_{\rm B} = \SI{-7}{K}$ with values given in Table~\ref{tab:Rvals}.
\begin{table}[h]
    \renewcommand{\arraystretch}{1.5}
    \setlength\tabcolsep{14pt}
    \begin{tabular}{@{}llll@{}} 
        \toprule
        $j$ & $R_{\rm min}^{(j)}$ [\si{\angstrom}] & $R_{\rm max}^{(j)}$ [\si{\angstrom}] & $\rho_{\rm rad}\qty(R_{\rm min}^{(j)})/\rho_{\rm rad}(0)$ \\
        \midrule
        0  & - & 0.0 & - \\
        1  & 1.7 & 3.2  & 0.0065\\
        2  & 4.6 & 6.2  & 0.012\\
        3  & 7.5 & 9.0  & 0.0049\\
        \bottomrule
    \end{tabular}
    \caption{\label{tab:Rvals} The locations of minima and maxima computed from the radial density in Figure~\ref{fig:radialDensity} at $\mu/\kB = \SI{-7}{\kelvin}$. The final column shows the vanishing density at the minima for the fully filled pore. These values are used in the computation of the linear density and coverage defined in Eqs.~(\ref{eq:rho1d})--(\ref{eq:rho2d}). The effects of temperature are negligible at the accuracy reported here.}
\end{table}
Temperature dependence of the minima/maxima locations only appear in the second digit not included in this table. The width of the central core $\approx \SI{3.4}{\angstrom}$ is only slightly larger than the van der Waals diameter of a helium atom (\SI{2.8}{\angstrom}) with over 95\% of atoms falling within this diameter when $\mu/\kB > \SI{-17}{\kelvin}$ at both temperatures studied.  Moreover, the final column of the table reports the radial density at the location of the first minima and shows it $100\times$ smaller than that in the center of the pore.  These combined results represents strong evidence for the quasi-1D nature of the central core.  

\begin{figure}[t]
    \centering
    \includegraphics[width=\columnwidth]{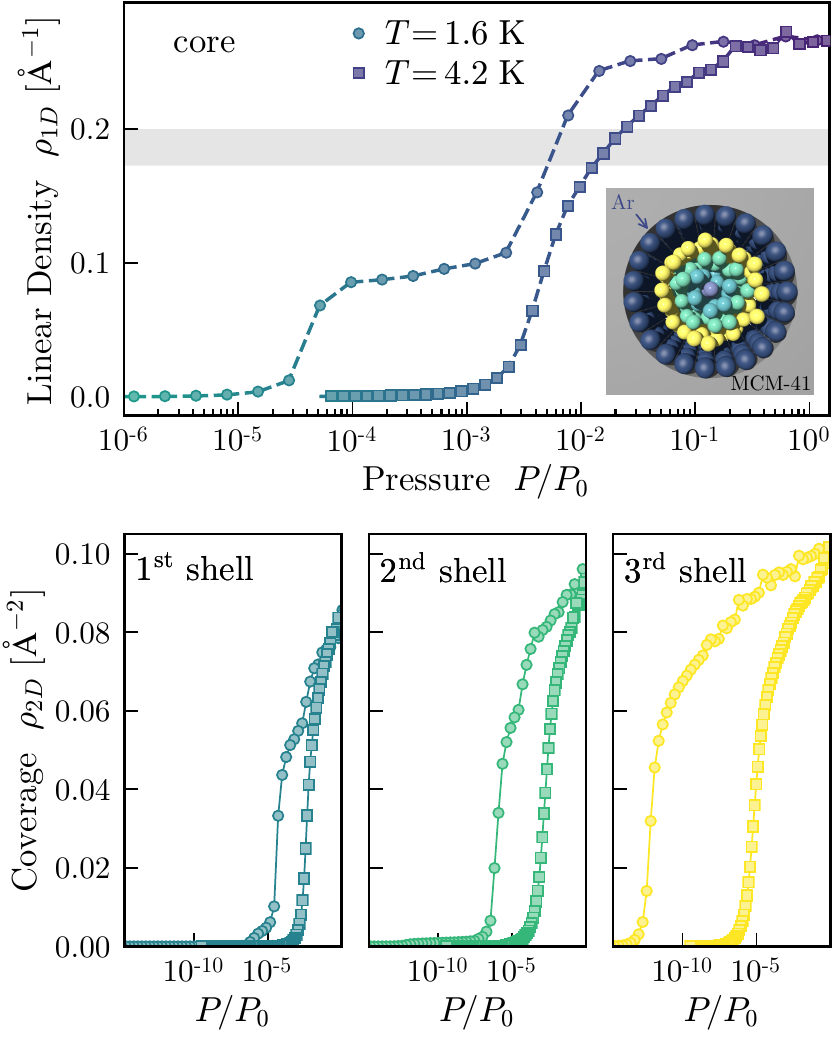}
    \caption{Top: The linear density $\rho_{\rm 1D}$ inside the central core as computed via Eq.~(\ref{eq:rho1d}) for two temperatures. The shaded bar indicates the range of pressures over which the quasi-1D helium has the same density as liquid $^4$He in the bulk.  A discussion of the step-like features is provided in the text. The inset shows a quantum Monte Carlo configuration of helium atoms inside the Ar-plated MCM-41 pore with distinct layers colored for emphasis. Bottom: The 2D area density or coverage $\rho_{2D}$ computed with Eq.~(\ref{eq:rho2d}) for the first, second, and third shell (as seen in the top panel inset) which are filled from outer-to-inner as the pressure is increased.}
\label{fig:linearDensity}
\end{figure}

Evaluating Eqs.~(\ref{eq:rho1d})--(\ref{eq:rho2d}) inside the pore we observe a strong pressure and temperature dependence of both the 1D linear and 2D coverage densities of the inner core and surrounding shells as seen in Figure~\ref{fig:linearDensity}. 
At $T=\SI{1.6}{\kelvin}$, the inner core begins to fill at lower pressures ($P \simeq 10^{-5}P_0$) and we first observe an increase in linear density which is due to the leakage of particles from the first shell as seen in the lower panel of Figure~\ref{fig:radialDensity}.  As the pressure is further increased, the density between the inner core and first shell begins to reduce and the quasi-1D nature of the central core becomes more apparent.  It is in this regime that we find a decade of pressures ($10^{-3}P_0 < P < 10^{-2} P_0$) where the inner core has a density equivalent to liquid $^4$He as indicated by the shaded bar in Figure~\ref{fig:linearDensity}.  Further increasing the pressure appears to quasi-solidify the core which is consistent with the expectation that atoms in 1D at high densities will seek to fix the distance between them to minimize their interaction potential $V(\vec{r})$. Results are similar at $T = \SI{4.2}{\kelvin}$ with the steps being smoothed out and the onset of the inner core being pushed to higher pressures due to the pressure dependence in Eq.~(\ref{eq:virialP}).  In the lower three panels, the areal coverage or 2D density of the surrounding cylinders is shown.  Here the onset is considerably sharper in pressure, with there being more tunability of the shell densities at higher temperature.

\section{Discussion}

The combination of experimental isotherms, neutron scattering, and X-ray scattering with large scale quantum Monte Carlo simulations provides unprecedented information on the structure of ${^4}$He inside small constrictions.  A BET analysis of experimental adsorption isotherms was used to determine the average pore radii of the grown MCM-41 ($\sim \SI{3}{\nano\meter}$) and the density of a single-layer of adsorbed Ar gas ($\SI{0.017}{\angstrom^{-3}}$).  These values were employed to build an effective model of the pre-plated adsorption potential which describes an infinite cylindrical cavity carved inside a continuous Lennard-Jones medium. While it is clear from the crystal structure and resulting full confinement potential shown in Figures~\ref{fig:mcm41} and~\ref{fig:manybodyPotential} that the atomic structure of bare MCM-41 produces a rough potential landscape, it is assumed that the pre-plating Ar gas will be first adsorbed into the deep potential pockets and resulting in a considerably smoothed environment.  This is justified by the agreement between experimental and theoretical adsorption isotherms seen in Figure~\ref{fig:filling_the_pore}. One of the main sources of uncertainty in this work is the conversion between the chemical potential (tuned in simulations) and the measurement of the partial pressure (controlled in the experiments).  It is expected that improvements could be made by implementing a direct Monte Carlo estimator for the pressure inside the nanopores combined with more sensitive determination of the pressure in the adsorption cell. 

As discussed in the introduction and validated in \S\ref{sec:experimental_results} and  \S\ref{sec:simulation_results}, in this low-pressure regime, $^4$He atoms are strongly attracted to the pre-plated walls of the MCM-41 constriction and a quasi-2d solid layer adsorbs on the surface, consistent with the scattering measurements at \SI{4.0}{\milli\mol\per\gram} and \SI{6.8}{\milli\mol\per\gram} in Figure~\ref{fig:scattering}.  The Ar layer has two main effects: (1) its weaker interaction with helium as compared to the MCM-41 reduces the potential well depth (this can be controlled via the density of the Ar layer); (2) the hard-cores of the Ar atoms decrease the effective radius of the constriction from \SI{15.5}{\angstrom} to $\sim \SI{11.75}{\angstrom}$.  We note that this is the exact reduction predicted from numerical simulations to ensure a cross-over from a quasi-3D to quasi-1D superfluid \cite{DelMaestro:2011dh,Kulchytskyy:2013dh,Markic:2020ip}.  Different confinement radii and potential environments could be obtained by changing the type of pre-plating rare gas, and these techniques could be utilized with other   
porous materials providing the opportunity to engineer a targeted confinement potential. 

As the partial pressure of helium is further increased, a series of concentric shells form separated by $r_m \simeq \SI{3}{\angstrom}$, the distance at which two helium atoms in free space would like to be situated to minimize their interaction energy. Shells near the Ar layer contain immobile helium, while those nearer the center of the pore may exhibit liquid-like behavior, consistent with the scattering results presented here.  The existence of finite density at the center of the pore is not guaranteed for a generic nanoporous system, even at large pressures \cite{DelMaestro:2012ba, Markic:2020ip}, and requires that the pre-plated pore radii is nearly commensurate with $r_m$. Again, numerical simulations have demonstrated that this central core can exhibit superfluid behavior \cite{Kulchytskyy:2013dh,Markic:2018bw,Markic:2020ip} at temperatures lower than those considered here.  

In conclusion, we have shown that theoretical simulations of $^4$He inside Ar pre-plated MCM-41 nanopores are in agreement with experimental adsorption and scattering results, and demonstrate the ability to stabilize a quasi-one-dimensional quantum liquid at low temperature with a tunable density at the center of the pores.   It is expected that such a 1D liquid should lack long-range order, instead displaying algebraically decaying density and phase correlations, even at zero temperature. This behavior could be confirmed in future work by further analyzing the results of experimental neutron scattering measurements along with quantum Monte Carlo calculations of the dynamical structure factor within the context of Tomonaga-Luttinger liquid theory.

\section{Acknowledgments}

We benefited from discussion with K. Shirahama and C. Herdman and acknowledge the support of the National Institute of Standards and Technology, U.S. Department of Commerce, in providing the neutron research facilities used in this work. 
This research was supported in part by the National Science Foundation (NSF) under award Nos.~DMR-1809027 and DMR-1808440.  Computations were performed on the Vermont Advanced Computing Core supported in part by NSF award No.~OAC-1827314 and IU Big Red Karst.  This work used the Extreme Science and Engineering Discovery Environment (XSEDE)\cite{xsede}, which is supported by NSF grant number ACI-1548562. XSEDE resources used include Bridges at Pittsburgh Supercomputing, Comet at San Diego Supercomputer Center, and Open Science Grid (OSG)\cite{osg1, osg2} through allocations TG-DMR190045 and TG-DMR190101.  OSG is supported by the NSF under award No.~1148698, and the U.S. Department of Energy's Office of Science.  Certain commercial equipment, instruments, or materials (or suppliers, or software, $\dots$) are identified in this paper to foster understanding. Such identification does not imply recommendation or endorsement by the National Institute of Standards and Technology, nor does it imply that the materials or equipment identified are necessarily the best available for the purpose.

\appendix
\section{Determination of Parameters for Effective MCM-41 Potential}
\label{app:potential}

Eq.~\eqref{eq:VHeMCM} in Section~\ref{ssec:preplatePotential} describes the full many-body potential experienced by a single $^4$He atom confined inside the MCM-41 matrix as a superposition of standard 6--12 Lennard-Jones interactions. Here we have used the optimized atomic coordinates available in Ref.~[\onlinecite{Ugliengo:2008ks}] and the potential was computed on a $\SI{40}{\angstrom} \times \SI{40}{\angstrom}$ pore-centered grid in the $xy$-plane at 101 different $z$ values from $z=\SI{-6.1}{\angstrom}$ to $z=\SI{6.1}{\angstrom}$.  At each of the grid points, we considered a semi-infinite crystal by summing over repeated unit cells until convergence was achieved at double floating point precision.

We performed an extensive search for the optimal Lennard-Jones parameters $\varepsilon_{ij}$ and $\sigma_{ij}$ to use in $U_{\mcm}(\vec{r_i})$ and found that a large range has been previously reported with minimal consensus in the literature. This included parameters based on Drieding force field calculations applied to silica, both with and without the effects of interacting hydrogen atoms
\cite{1990:MayoJPhysChem, 2008:ZhuoJPhysChemC, Jing:2013im,
Chang:2015jd, 2018:KumarKoreanJChemEng,2011:HoLangmuir}, and including
a large variation of the interaction strength ($\SI{75}{\kelvin} \leq
\varepsilon/\kB \leq \SI{230}{\kelvin}$) of both oxygen and silicon in the MCM-41 
\cite{Williams:2016eq, 1996:BrodkaJChemPhys,
2014:Ferreiro-RangelJPhysChemC, 2006:SchumacherJPhysChemB, 2001:TaluColloidsSurf,
2016:ZhangLangmuir}.  We have chosen to employ an alternative set of parameters that derived from consistent valence force field calculations \cite{2011:CoasneJPhysChemC, 1994:HillJPhysChem} that have been designed specifically for silicates and have been able to reproduce experimentally observed crystal structures.  The appropriately mixed parameters used in the simulations are reported in Table~\ref{tab:LJParams}. 
\begin{table}[h]
    \renewcommand{\arraystretch}{1.5}
    \setlength\tabcolsep{12pt}
    \begin{tabular}{@{}lll@{}} 
        \toprule
        Atom Pair & $\sigma_{ij}\; \qty[\si{\angstrom}]$ & $\varepsilon_{ij}/k_\mathrm{B}\; \qty[\si{\kelvin}]$ \\
        \midrule
        Si--He     & 3.60 & 14.84 \\
        O--He      & 2.93 & 35.44 \\
        H--He      & 2.70 & 12.13 \\
        \bottomrule
    \end{tabular}
    \caption{\label{tab:LJParams} Lennard-Jones potential parameters
    for interactions between helium and each atom within the MCM-41 crystal structure \cite{2011:CoasneJPhysChemC,1964:HirschfelderBook}. Values were determined
    via standard Lorenz-Bertholot mixing rules \cite{Boda:2008yy}, Eq.~(\ref{eq:LBrules}).}
\end{table}

\begin{figure}[t]
\includegraphics[width=\columnwidth]{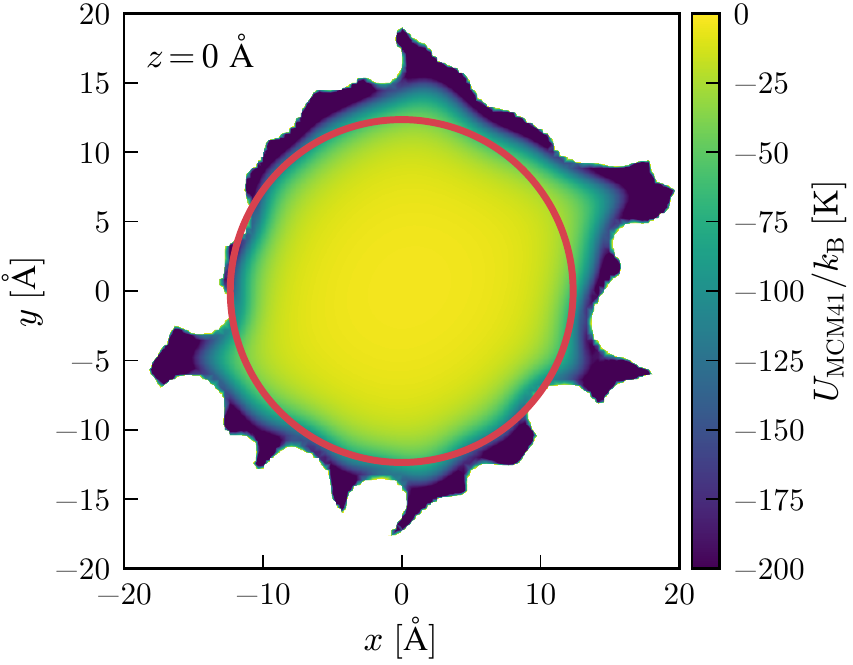}
\caption{The confinement potential data at $z=\SI{0.0}{\angstrom}$ with $U_{\mcm}/\kB \le \SI{0.0}{\kelvin}$ over which a fit was performed to $U_{\rm cyl}$ in Eq.~\eqref{eq:Vpore}. The potential cutoff method uses all the data displayed, while the radial cutoff uses only those data within the red circle of radius \SI{12.4}{\angstrom}. A minimum cutoff value for the potential of $\SI{-200}{\kelvin}$ has been used here for display purposes.} 
\label{fig:cylindrical_potential_effective_parameters_z0}
\end{figure}

With the full potential calculated, we next obtained the effective potential $U_{\rm cyl}$ in Eq.~\eqref{eq:Vpore} to be employed in grand canonical quantum Monte Carlo simulations via a non-linear least squares fitting procedure at each value of $z$.  This is made difficult by the infinitely repulsive pore walls and the proximate deep attractive minima.  To overcome these extremal values two alternative methods were employed to obtain a well-behaved dataset over which to fit:  a \emph{potential cutoff} selected spatial grid-points inside the MCM-41 pore where $U_{\mcm}/\kB\leq\SI{0.0}{\kelvin}$,  and a \emph{spatial cutoff} where all points within a fixed radius set by the closest point to the center with $U_{\mcm}/\kB=\SI{0.0}{\kelvin}$ were chosen. The resulting spatial regions over which fits were performed at $z=\SI{0.0}{\angstrom}$ are compared in Figure~\ref{fig:cylindrical_potential_effective_parameters_z0}.
The potential cutoff method more accurately captures the pore roughness as seen in the figure, while the radial cutoff method provides an improved description of the region of the pore ($r < \SI{12.4}{\angstrom}$) where the helium atoms will reside after the pre-plating has occurred. 

\begin{figure}[t]
    \centering
    \includegraphics[width=\columnwidth]{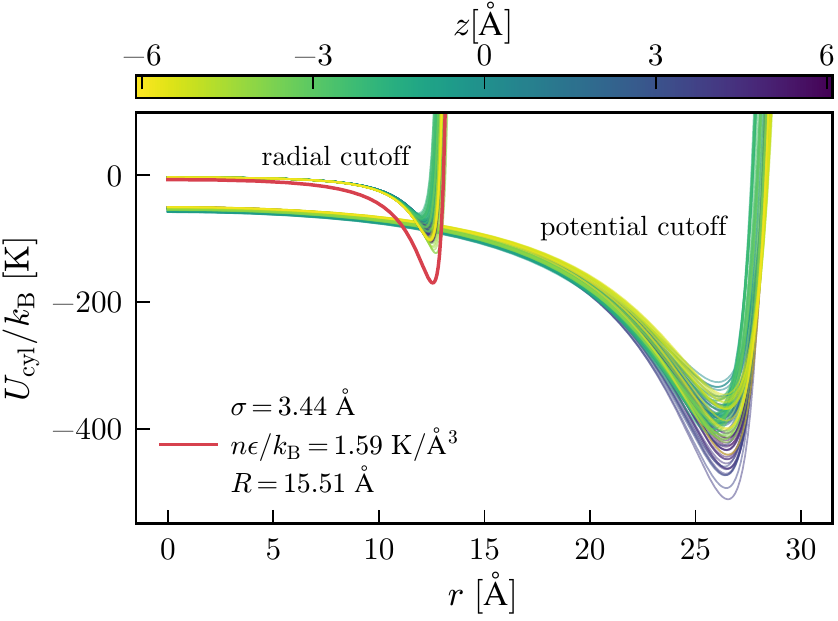}
    \caption{Comparison of two different methods for fitting Eq.~(\ref{eq:Vpore}) to Eq.~(\ref{eq:VHeMCM}) with different lines corresponding to potential slices within the $xy$-plane from $z=\SI{-6.1}{\angstrom}$ to $z=\SI{6.1}{\angstrom}$ along the pore.  The thick red line corresponds to the displayed parameters used for this study.}
    \label{fig:cylindrical_potential_effective_parameters}
\end{figure}

The fitting procedure from $U_{\mcm}$ to $U_{\rm cyl}$ then proceeds by determining the optimal set of parameters $\sigma, n\epsilon$, and $R$ in Eq.~\eqref{eq:Vpore} at each of the 101 $z$-slices between $z = \SIrange{-6.1}{6.1}{\angstrom}$. The resulting smooth symmetric confinement potentials for both methods are shown in Figure~\ref{fig:cylindrical_potential_effective_parameters}. Upon comparison, it is clear that the raw potential cutoff does not provide a reasonable effective potential as it predicts a confinement radius that is inconsistent with experimental adsorption isotherms.  This is due to the fact that it is over-fitting to the large potential well depths within the rough areas of the pore and neglecting the smoother lower potential region closer to the center where the $^4$He will be confined. As mentioned above, upon pre-plating we expect these volumes to be filled by Ar and result in a considerably smoother potential with a smaller effective radius.  This is better-captured by the radial cutoff method which also produces an effective confinement radius that is in excellent agreement with the experimentally determined pore diameter.  We thus focus on the results of the radial cutoff approach and determine the final set of parameters by performing a weighted average over the $z$-slices and choosing a value of $n\epsilon$ that incorporates the expected added density of argon due to its filling the extremal regions causing a deeper potential well.  The resulting effective potential $U_{\rm cyl}$ is shown as a red solid line in Figure~\ref{fig:cylindrical_potential_effective_parameters_z0} along with the parameters, which are also reported in Table~\ref{tab:MCM41}. We note that there is considerable ambiguity in this procedure and appeal to the reasonable agreement found between theoretical and experimental results reported in \S\ref{sec:simulation_results}.  A more microscopic characterization of the helium--MCM-41 interaction combing classical grand canonical Monte Carlo with molecular dynamics simulations is ongoing, but is beyond the scope of this study.

\section{Simulation Scaling}

\subsection{Finite Size Effects}
\label{ssec:finite_size_scaling}

\begin{figure}[t!]
    \centering
    \includegraphics[width=\columnwidth]{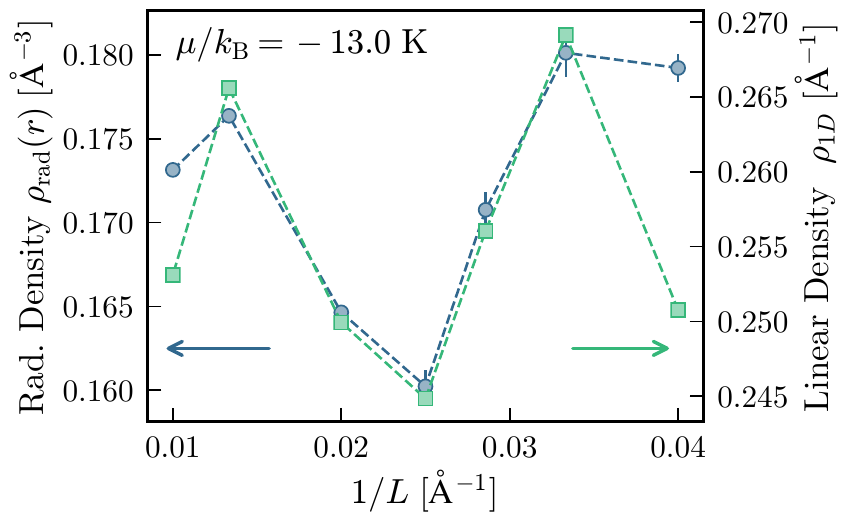}
    \caption{The volume ($\rho_{\rm rad}(r=0)$, circles) and linear ($\rho_{\rm 1D}$, squares) density of the 1D core inside a Ar plated MCM-41 nanopore shows strong finite size effects at fixed temperature and chemical potential. The non-monotonic behavior is consistent with decaying oscillations arising for the interplay between pore length and the average separation between particles in the core.}
    \label{fig:finite_size_scaling_inner_core}
\end{figure}

To understand the effects of finite pore length, we performed grand canonical quantum path integral Monte Carlo simulations for helium confined within argon pre-plated MCM-41 with $L=\SIrange{25.0}{100}{\angstrom}$. Maximal effects were observed when the pore was completely filled with $\mu/\kB=\SI{-13.0}{\kelvin}$ at $T=\SI{1.6}{\kelvin}$.  
\begin{figure}[t!]
    \centering
    \includegraphics[width=\columnwidth]{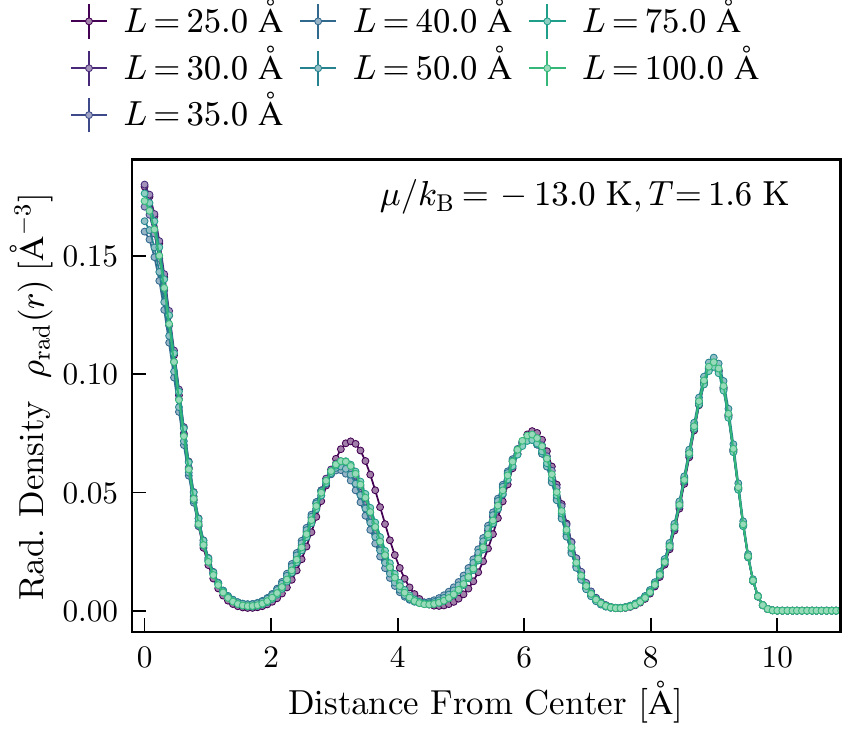}
    \caption{Structural finite size effects in the radial volume density of different pore lengths $L$ in grand canonical quantum Monte Carlo simulations of $^4$He confined inside Ar pre-plated MCM-41 at fixed temperature and chemical potential.  For $L \ge \SI{50}{\angstrom}$, the location of adsorbed shells (peaks) and the minima between layers are mostly unaffected by the finite pore length.}
    \label{fig:finite_size_scaling_radial_density}
\end{figure}
Figure~\ref{fig:finite_size_scaling_radial_density} shows that there is little structural change, as captured by the radial density with only minimal shifts of the peak densities and  minima between layers.  As we are most interested in the $^4$He confined at the very center of the pore, we can quantify these effects by studying the length dependence of $\rho_{\rm rad}(r = 0)$ as well as the integrated linear density $\rho_{\rm 1D}$ defined in Eq.~\eqref{eq:rho1d} as seen in Figure~\ref{fig:finite_size_scaling_inner_core} for the same set of simulation parameters.  On this reduced scale, the finite size effects are more striking and exhibit non-monotonic oscillatory behavior.  
This can be understood as the confluence of different but related finite size effects. For $L=\SI{25}{\angstrom}$, the aspect ratio of the pore is $\sim 1:1$ and its length can accommodate approximately $6$ helium atoms at its center. Such a small system can not be expected to capture the physics in the thermodynamic limit.  As $L$ is increased, there are oscillations caused by the interplay between the length and the optimal inter-particle separation $r_m \simeq \SI{3}{\angstrom}$ set by the He-He interaction as only an integer number of atoms can fit in the center of the pore. This effect becomes less important for longer pores leading to a decay in the observed density oscillations.  Finally, the total density of the ${}^4$He atoms inside the pore comes with its own finite size errors due to the $1/N$ scaling of the chemical potential $\mu$ (see Ref.~[\onlinecite{Herdman:2014bx}])  which has not been considered here. Correcting for this would require the comparison of simulations performed at different length-dependent chemical potentials.

\subsection{Trotter Error}
\label{ssec:tau_scaling}

\begin{figure}[h]
    \centering
    \includegraphics[width=\columnwidth]{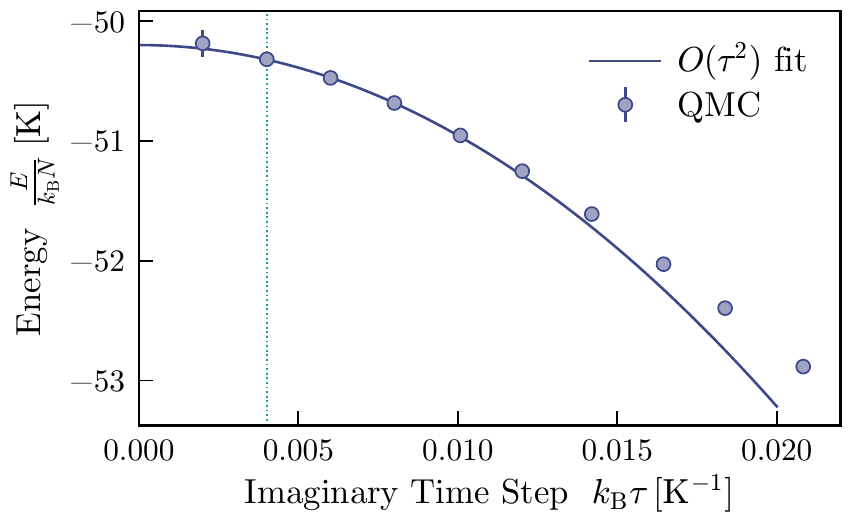}
    \caption{The effects of the finite imaginary time step $\tau$ on the energy per particle of $^4$He confined inside Ar pre-plated MCM-41 pores.  Data points are from quantum Monte Carlo (QMC) simulations at $T = \SI{1.6}{\kelvin}$,  $L = \SI{2.5}{\nano\meter}$, and $\mu/\kB=\SI{-50}{\kelvin}$. The solid line is a fit to $E/N = E/N\rvert_{\tau \to 0} + \mathcal{C}\tau^2$ and the dotted vertical line shows the value $\kB\tau = \SI{0.004}{\kelvin^{-1}}$ that was used in all production simulations.}
    \label{fig:trotter_error}
\end{figure}

Path integral quantum Monte Carlo is stochastically exact at finite
temperature, up to a systematic finite-time step (Trotter) error that can be
controlled through an appropriate implementation of a composite factorization
scheme for the density matrix \cite{Jang:2001cl}. Here we use a
$\mathcal{O}(\tau^4)$ approach for the He-He interaction, but are limited to a
primitive $\mathcal{O}(\tau^2)$ approximation for the confinement potential. The
resulting effect on simulation results can be seen in
Figure~\ref{fig:trotter_error}, which shows the energy per particle of $^4$He
inside Ar pre-plated MCM-41 as a function of imaginary time step $\tau$ at $T =
\SI{1.6}{\kelvin}$.  The simulation data is well fit by a quadratic polynomial
in $\tau$ as shown by the solid line.  The extrapolation of $\tau \to 0$ is
within 1\% of the value at  $\tau = \SI{1/250}{\kelvin^{-1}}$ (vertical
dotted line) that we have utilized for all reported results to balance accuracy
with simulation efficiency. 

\FloatBarrier

\nocite{apsrev41Control}
\bibliographystyle{apsrev4-1}
\bibliography{refs}

\end{document}